\documentclass[prd,12pt,nofootinbib,superscriptaddress]{revtex4-2}
\usepackage{bm}
\usepackage{amsmath}
\usepackage{amssymb}
\usepackage{graphicx}
\usepackage{subfigure}
\usepackage{hyperref}
\usepackage{color}
\usepackage{CJK}
\usepackage{orcidlink}
\usepackage{array}
\usepackage{indentfirst}

\hypersetup{
	colorlinks=true,
	linkcolor=red,
	citecolor=blue,
}

\begin{document}

\begin{CJK*}{UTF8}{gbsn}

\title{Generic effective sources for first-order in mass-ratio gravitational self-force calculations in Schwarzschild spacetime}

\author{Chao Zhang\orcidlink{0000-0001-8829-1591}}
\email{zhangchao1@nbu.edu.cn}
\affiliation{Institute of Fundamental Physics and Quantum Technology, Department of Physics, School of Physical Science and Technology, Ningbo University, Ningbo, Zhejiang 315211, China}

\author{Rong-gen Cai\orcidlink{0000-0002-3539-7103}}
\email{caironggen@nbu.edu.cn}
\affiliation{Institute of Fundamental Physics and Quantum Technology, Department of Physics, School of Physical Science and Technology, Ningbo University, Ningbo, Zhejiang 315211, China}

\author{Guoyang Fu\orcidlink{0000-0002-7944-8611}}
\email{fuguoyangedu@sjtu.edu.cn}
\affiliation{School of Physics and Astronomy, Shanghai Jiao Tong University,  Shanghai 200240, China}

\author{Yungui Gong\orcidlink{0000-0001-5065-2259}}
\email{Corresponding authors. gongyungui@nbu.edu.cn}
\affiliation{Institute of Fundamental Physics and Quantum Technology, Department of Physics, School of Physical Science and Technology, Ningbo University, Ningbo, Zhejiang 315211, China}

\author{Xuchen Lu\orcidlink{0000-0002-9093-9059}}
\email{luxuchen@nbu.edu.cn}
\affiliation{Institute of Fundamental Physics and Quantum Technology, Department of Physics, School of Physical Science and Technology, Ningbo University, Ningbo, Zhejiang 315211, China}

\author{Wenting Zhou\orcidlink{0000-0003-4046-753X}}
\email{zhouwenting@nbu.edu.cn}
\affiliation{Institute of Fundamental Physics and Quantum Technology, Department of Physics, School of Physical Science and Technology, Ningbo University, Ningbo, Zhejiang 315211, China}

\date{\today}

\begin{abstract}
The numerical calculation of gravitational self-force in extreme mass ratio inspiral systems is fundamentally challenging due to the singular nature of point-particle sources.
To overcome these difficulties, the effective source method offers an innovative alternative by replacing traditional regularization techniques with a reformulation of the problem.
In this paper, we present the first fully analytic framework for constructing effective sources to compute the gravitational self-force for generic orbits in Schwarzschild spacetime.
By reformulating the singular field through angular modulation in terms of a tetrad decomposition, the effective source can be constructed with the linear combination of scalar modes.
The derived effective source is continuous across the particle's worldline, enabling efficient numerical implementation in $1+1$ dimensions.
\end{abstract}

\maketitle

\end{CJK*}

\section{Introduction}
Extreme mass ratio inspirals (EMRIs), which compact objects like neutron stars or stellar-mass black holes are captured by a supermassive black hole (SMBH)  with masses in the range of $10^6-10^9~M_{\odot}$ in galactic nuclei  \cite{Magorrian:1997hw,Ghez:2008ms,Gillessen:2008qv,EventHorizonTelescope:2022wkp}, are the most scientifically valuable targets for space-based detectors such as the Laser Interferometer Space Antenna (LISA) \cite{Danzmann:1997hm,LISA:2017pwj}, TianQin \cite{Luo:2015ght}, and Taiji \cite{Hu:2017mde}.
In a typical EMRIs system, the small compact object undergoes orbital inspiral for at least several years while executing $10^5-10^6$ detectable cycles in a closely bound orbit around the SMBH prior to merger \cite{Barack:2018yvs}.
The unique information encoded in EMRIs gravitational wave (GW) signals will provide fundamental insights into black hole physics \cite{Barsanti:2022vvl, Guo:2022euk, Maselli:2021men, Maselli:2020zgv, Barsanti:2022ana, Yunes:2011aa, Cardoso:2011xi, Zhang:2022rfr,AbhishekChowdhuri:2022ora,Torres:2020fye, Zhang:2022hbt, Zhang:2023vok, Liang:2022gdk,Bhattacharyya:2023kbh,Ghosh:2024arw}, dark matter environments in galactic nuclei \cite{Becker:2022wlo, Cardoso:2019rou, Kavanagh:2020cfn, Eda:2014kra, Li:2021pxf, Dai:2021olt,AbhishekChowdhuri:2023cle,Barausse:2014tra,Barausse:2014pra,Speri:2022upm,Aghaie:2023lan,Cardoso:2022whc}, and the nonlinear regime of gravity \cite{Barausse:2014tra,Centrella:2010mx,Lehner:2014asa}.
The signals are obscured by detector noise, making their identification possible only through matched filtering techniques using precise GW templates.
The detection and analysis of GWs from EMRI systems critically depends on the availability of accurate theoretical waveform templates.
Numerical relativity has achieved remarkable progress in simulating the merger of stellar-mass compact binary systems \cite{Pretorius:2005gq,Fedrow:2017dpk,Gourgoulhon:2007ue}.
By solving Einstein's field equations, numerical relativity has successfully modeled the millisecond-scale dynamics during the merger of such systems, laying a solid foundation for the first discovery of GW event GW150914 detected by the Laser Interferometer Gravitational-Wave Observatory (LIGO) Scientific Collaboration and the Virgo Collaboration \cite{LIGOScientific:2016emj}.
However, for EMRIs system with orbital evolution spanning months or even years,
numerical relativity requires prohibitively large computational resources, rendering the task nearly impossible.
Meanwhile, post-Newtonian methods are only applicable in weak-field and low-velocity regimes, making them unsuitable for the strong-field and relativistic motion involved in EMRI systems \cite{Einstein:1938yz,Chandrasekhar:1965gcg}.
These inherent fundamental limitations have led to the development of the gravitational self-force (GSF) formalism as a uniquely powerful approach specifically tailored for systems with EMRIs \cite{Detweiler:2008ft,Blanchet:2009sd,Wardell:2024yoi,Leather:2024mls,Hollands:2024iqp,Dolan:2023enf,Bourg:2024vre,NavarroAlbalat:2022tvh,Detweiler:2002gi,Wardell:2015kea}.

The GSF formalism provides a particularly powerful framework for modeling EMRIs with no restriction about orbital velocity during the entire evolution \cite{Mino:1996nk,Quinn:1996am}.
The method employs a systematic perturbative expansion in the small parameter of the mass ratio.
In the zeroth-order mass-ratio expansion, the small object  traces a geodesic trajectory uniquely determined by the background spacetime of the supermassive black hole.
The first-order mass-ratio correction brings in the vital GSF effects, 
where the small body's gravitational field exerts feedback on the geodesic motion, 
leading to gradual deviations from geodesic trajectories.
Meanwhile, The applicability of the GSF formalism can be extended to systems with intermediate mass ratios through higher-order mass-ratio expansions, thereby bridging the gap with numerical relativity techniques for such intermediate mass ratio systems \cite{Wardell:2021fyy,Albertini:2022rfe,Burke:2023lno}.

The major obstacle to the implementation of the GSF formalism is the singularity of the gravitational field at the particle's position when the smaller compact object is regarded as a mere point particle.
To circumvent the singularity, regularization techniques must be employed to extract meaningful and finite results from the divergent gravitational perturbation field.
The physical field can be decomposed into singular and regular components that the singular part contains the point-particle singularity but exerts no force on the particle while the regular part contributes the force that completely governs the orbital dynamics \cite{Poisson:2003nc,Mino:1996nk}.
Detweiler and Whiting gives the local expansion of the singular field  via a Hadamard form  near the worldline \cite{Detweiler:2002mi,Heffernan:2012su}.
The decomposition provides the foundation for practical regularization schemes, such as the mode-sum method \cite{Barack:1999wf,Barack:2010tm,vandeMeent:2017bcc} and the effective-source approach \cite{Vega:2007mc,Barack:2007jh,Vega:2011wf}.

Up to now, most of the first-order mass-ratio GSF calculations have been carried out by confining the particle's motion to a geodesic of the background spacetime.
Modeling the orbital evolution as a smooth sequence of geodesics does not account for the conservative effect of GSF.
In order to incorporate the dissipative and conservative effect of GSF, the idea of osculating geodesics was developed \cite{Pound:2007th,Barack:2018yvs}.
This method reconstructs the inspiral trajectory from a continuous family of geodesics, where each member is tangent to the true worldline at a specific instant.
At each instant $t_0$, the GSF is approximated by calculating it for the geodesic that is tangent to the evolving orbit at that instant, treating this geodesic as the particle's past trajectory.
Additionally, multiscale expansion where the full accelerated worldline is expanded around another accelerated worldline has been applied to the second-order self-force calculations \cite{Miller:2020bft,Pound:2019lzj,Warburton:2021kwk,Wardell:2021fyy}.
The only fundamental error inherent in these procedures arises from the reliance on geodesic-based GSF data, which is calculated under the approximation of a fixed orbit instead of the true dynamical trajectory.
Quantifying the phase error introduced by the geodesic self-force approximation requires comparison with self-consistent calculations performed via direct time-domain evolution.
In such evolution, the GSF is computed at each time step, and the orbit sourcing the field is advanced accordingly using the updated force.
So far, in these evaluations, only the case of a scalar-charged particle has been considered \cite{Heffernan:2017cad,Diener:2011cc}.
Furthermore, a rough estimate suggests that second-order mass-ratio self-force must be taken into account in waveform modeling to achieve phase accuracy at the order of unity for matched filtering \cite{Hinderer:2008dm,Miller:2020bft},
whereas the mode-sum method is only applicable to first-order mass-ratio self-force calculations.
Prior work on scalar self-force calculations demonstrated the efficacy of the effective source method for the generic orbits \cite{Leather:2023dzj,Wardell:2011gb,Bourg:2024cgh,Dolan:2010mt} as well as for the self-consistent orbits \cite{Diener:2011cc}.

The effective source approach reformulates the Einstein equations with a finite and regularized source term, particularly through covariant expansions of the Detweiler-Whiting singular field \cite{Vega:2007mc,Barack:2007jh,Vega:2011wf}.
The formally divergent GSF problem can be transformed into a mathematically well-defined and computationally tractable framework \cite{Vega:2007mc}.
During the numerical evolution, the continuous solutions for the regularized metric perturbation are directly yielded and offer significant advantages for various aspects of GSF calculations especially on self-consistent GSF and second-order mass-ratio GSF \cite{Wei:2025lva,Upton:2023tcv,Upton:2021oxf,Spiers:2023mor,Miller:2023ers}.
Currently, the effective source method has only been applied to circular orbit scenarios for GSF, while calculations for generic orbits and self-consistent orbital evolution remain to be completed \cite{Wardell:2015ada}.
One of the most crucial difficulties for extending scalar self-consistent results to the gravitational case is constructing an effective source that allows for rapid and accurate calculations \cite{Heffernan:2017cad}.
A central objective of this work is to construct the generic effective source of gravitational field suitable for the self-consistent and second-order mass-ratio inspiral evolution in Schwarzschild spacetime.

The paper is organized as follows.
In Sec. \ref{sec2}, we briefly introduce the formalism of the effective source approach and derive the required second-order-distance singular field for constructing the continuous and finite effective source.
In Sec. \ref{sec4}, the tetrad decompositon method is developed to derive the analytical effective source formula for generic orbit in terms of the scalar harmonics.
In Sec. \ref{sec6}, we give example calculations of effective source for the cases of a generic orbit in Schwarzschild spacetime.
Section \ref{sec7} is devoted to conclusions and discussions.

\section{Effective Source Approach and The Singular Field}\label{sec2}

Introducing the trace-reversed perturbation $\bar{h}_{\mu\nu} \equiv h_{\mu\nu} - g_{\mu\nu}h/2$,
and imposing the Lorenz gauge condition $\nabla_\mu \bar{h}^{\mu\nu}=0$,
the linearized Einstein equations for the perturbation $\bar{h}_{\mu\nu}$ are
\begin{equation}\label{lEQ}
\square\bar{h}_{\mu\nu} + 2 R^{\rho\;\;\sigma}_{\;\;\mu\;\;\nu}\bar{h}_{\rho\sigma} = -16\pi T_{\mu\nu}\, ,
\end{equation}
where the covariant derivative $\nabla_\mu$ and the Riemann tensor $R^{\rho\;\;\sigma}_{\;\;\mu\;\;\nu}$ are computed  with respect to the background metric $g_{\mu\nu}$,
the energy-momentum tensor of the point particle is given by
\begin{equation}\label{eq:energytensor}
	T_{\mu\nu} = \mu\int^\infty_{-\infty} [-\det(g)]^{-1/2} \delta^4(x^\mu - x_p^\mu(\tau))u_\mu u_\nu\, d\tau ,
\end{equation}
$\mu$ is the mass of the small compact object modeled as a point particle,  $\det(g)$ is the determinant of the background metric, $u^\mu$ denotes the particle's four-velocity, and $\tau$ denotes the proper time along the particle's worldline.

The numerical solution of the field equations \eqref{lEQ} and \eqref{eq:energytensor} is complicated by the singular nature of the point-particle source.
To overcome this difficulty, a common approach is to isolate the meaningful, finite part of the perturbation by decomposing the perturbation into singular and regular components,
\begin{equation}\label{eq:hsr}
\bar{h}_{\mu\nu}=\bar{h}^{R}_{\mu\nu}+\bar{h}^{S}_{\mu\nu},
\end{equation}
where the regular part $\bar{h}^{R}_{\mu\nu}$
is finite and physically meaningful for numerical computations such as for evaluating the GSF,
while the singular part $\bar{h}^{S}_{\mu\nu}$
captures the divergent structure analytically.
Substituting Eq. \eqref{eq:hsr} into the linearized Einstein equation \eqref{lEQ} yields the reformulated equation for the regular part,
\begin{equation}
\square\bar{h}^R_{\mu\nu} + 2 R^{\rho\;\;\sigma}_{\;\;\mu\;\;\nu}\bar{h}^R_{\rho\sigma} = -16\pi T_{\mu\nu} -\square\bar{h}^S_{\mu\nu} - 2 R^{\rho\;\;\sigma}_{\;\;\mu\;\;\nu}\bar{h}^S_{\rho\sigma}.
\end{equation}
The Detweiler-Whiting singular field $\bar{h}_{\mu\nu}^S$
can be derived analytically only near the particle's worldline and is typically truncated at a finite-order distance;
this truncated approximation is known as the puncture field $\bar{h}_{\mu\nu}^P$.

To ensure that the effective source is both continuous and finite,
the required order of the puncture function varies significantly depending on the computational scheme employed. Previous studies have shown that in
$1+1$D schemes, a second-order-distance puncture is sufficient to produce an effective source that is finite and continuous at the particle's worldline \cite{Barack:2007we}.
In contrast, full $3+1$D frameworks require at least a fourth-order distance-puncture construction to achieve the same level of regularity in the effective source \cite{Wardell:2011gb}.
In this work, we develop a rigorous framework for constructing the effective source within $1+1$D numerical schemes for GSF calculations.
The covariant expansion of the Detweiler-Whiting singular field in Schwarzschild spacetime is given by \cite{Heffernan:2012su}
\begin{equation}\label{SingularField}
  \bar{h}^S_{\mu\nu}=
4 \mu g_{\mu}^{~\bar{\mu}} g_\nu^{~\bar{\nu}} \Big\{\frac{1}{\eta}
    \frac{u_{\bar{\mu}} u_{\bar{\nu}}}{\bar{s}} +\mathcal{O}(\eta)\Big\},
\end{equation}
where
\begin{equation}
\bar{s}^2 \equiv (g_{\bar{\mu} \bar{\nu}} + u_{\bar{\mu}} u_{\bar{\nu}}) \sigma^{\bar{\mu}} \sigma^{\bar{\nu}},
\end{equation}
$g_{\bar{\mu} \bar{\nu}}$ and $u^{\bar{\mu}}$ are the background metric and four-velocity evaluated at the particle position,
the parameter $\eta$ counts the order of small distance between the field point and the particle position,
the gradient of Synge's function $\sigma_{\bar{\alpha}}$, the parallel propagator $g_{\alpha}^{~\bar{\mu}}$ are \cite{Heffernan:2012su}
\begin{equation}\label{sigmagradient}
-\sigma_{\bar{\alpha}}(x,x_p)=g_{\alpha\beta}w^{\beta}+A_{\alpha\beta\gamma}w^{\beta}w^{\gamma}+A_{\alpha\beta\gamma\delta}w^{\beta}w^{\gamma}w^{\delta}+A_{\alpha\beta\gamma\delta\epsilon}w^{\beta}w^{\gamma}w^{\delta}w^{\epsilon}+\cdots,
\end{equation}
\begin{equation}
g_{\alpha}^{~\bar{\mu}}(x,x_p)=\delta_{\alpha}^{\mu}+B_{\alpha\beta}^{\mu}w^{\beta}+B_{\alpha\beta\gamma}^{\mu}w^{\beta}w^{\gamma}+B_{\alpha\beta\gamma\delta}^{\mu}w^{\beta}w^{\gamma}w^{\delta}+\cdots.
\end{equation}
Here all coefficients are evaluated at the particle position $x_p$, and the coordinate difference between the field point $x=[t_p, r, \theta, \phi]$ and particle position $x_p=[t_p, r_p, \frac{\pi}{2}, \phi_p]$ is
\begin{equation}
w^\alpha=x^\alpha-x_p^\alpha=[0,\delta r,\delta\theta, \delta\phi].
\end{equation}
After extensive symbolic manipulation, we obtain the second-order-distance puncture field in the form
\begin{equation}\label{hpfunction}
\bar{h}_{\mu\nu}^P= \frac{4\mu \chi_{\mu\nu}}{\bar{s}},
\end{equation}
where the nonvanishing components of $\chi_{\mu\nu}$ are
\begin{equation}
\begin{split}
\chi_{tt}&=u_tu_t+\frac{2u_tu_t}{r_p(r_p-2)}w^1,\qquad \chi_{tr}=u_tu_r +\frac{u_tu_\phi }{r_p}w^3,\\
\chi_{t\theta}&=-(r_p-2)u_tu_r w^2,\\
\chi_{t\phi}&=-(r_p-2)u_tu_r w^3+u_tu_\phi +\frac{(r_p-1)u_tu_\phi }{(r_p-2)r_p}w^1,\\
\chi_{rr}&=u_r u_r -\frac{2u_r u_r }{(r_p-2)r_p}w^1+\frac{2u_r u_\phi }{r_p}w^3,\\
\chi_{r\theta}&=-(r_p-2)u_r u_r w^2,\\
\chi_{r\phi}&=u_r u_\phi +\frac{(r_p-3)u_r u_\phi }{(r_p-2)r_p}w^1+\left(\frac{u_\phi u_\phi }{r_p}-(r_p-2)u_r u_r \right)w^3,\\
\chi_{\theta\phi}&=-(r_p-2)u_r u_\phi w^2,\\
\chi_{\phi\phi}&=-2(r_p-2)u_r u_\phi w^3+u_\phi u_\phi +\frac{2u_\phi u_\phi }{r_p}w^1.
\end{split}
\end{equation}
To correct for the discontinuity at $\delta\phi=\pm\pi$ in the puncture function defined by Eq.~\eqref{hpfunction},
we replace the term $\delta\phi$ with \cite{Dolan:2012jg,Heffernan:2017cad,Wardell:2015ada}
\begin{equation}
\delta\phi\to \sin \delta\phi~[=\delta\phi+\mathcal{O}(\delta\phi^3)],
\end{equation}
which preserves the local expansion of $\bar{h}_{\mu\nu}^P$
near the particle and keeps it consistent with the singular field through second-order distance.
Similarly, the term $\delta\theta$ is replaced by
\begin{equation}
\delta\theta\to \sin \delta\theta~[=\delta\theta+\mathcal{O}(\delta\theta^3)].
\end{equation}

\section{Spherical-Harmonic decomposition}\label{sec4}
Owing to the spherical symmetry of the background geometry,
the metric perturbation can be decomposed using Barack-Lousto-Sago (BLS) tensor spherical harmonics \cite{Barack:2005nr,Barack:2007tm},
\begin{equation}\label{halphabeta}
\bar{h}_{\alpha\beta}=\frac{\mu}{r}\sum_{l=0}^{\infty}\sum_{m=-l}^{l}\sum_{i=1}^{10}a_{l}^{(i)}\bar{h}^{(i)lm}(r,t)Y_{\alpha\beta}^{(i)lm}(\theta,\varphi;r),
\end{equation}
where the normalization factors are given by
\begin{equation}
a^{(i)}_l=\frac{1}{\sqrt{2}}\times\left\{\begin{array}{ll}
1, & i=1,2,3,6,        \\
{[}l(l+1)]^{-1/2},  & i=4,5,8,9, \\
{[}l(l-1)(l+1)(l+2)]^{-1/2}.  & i=7,10,
\end{array}
\right.
\end{equation}
the BLS tensor spherical harmonics $Y_{\alpha\beta}^{(i)lm}$ are given in  \cite{Barack:2005nr,Barack:2007tm}.
The $Y_{\alpha\beta}^{(i)lm}(\theta,\varphi;r)$ constitute an orthonormal basis
\begin{equation}
\int \tilde{\eta}^{\alpha\mu}\tilde{\eta}^{\beta\nu}[Y_{\mu\nu}^{(i)lm}]^{*}Y_{\alpha\beta}^{(j)l^{\prime}m^{\prime}} d\Omega=\delta_{ij}\delta_{ll^{\prime}}\delta_{mm^{\prime}}
\end{equation}
where $\tilde{\eta}^{\alpha\gamma}=\mathrm{diag}[1,f^2,r^{-2},r^{-2}\sin^{-2}\theta]$.
We substitute Eq.~\eqref{halphabeta} into the linearized Einstein equation \eqref{lEQ}
and can get the coupled set of partial differential equations \cite{Barack:2005nr}
\begin{equation}
\square_{2d}\overline{h}^{(i)lm}+\mathcal{M}^{(i)l}_{\;(j)}\overline{h}^{(j)lm}=T^{(i)lm}\delta(r-r_p)\quad(i=1,\ldots,10),
\end{equation}
where the wave operator is
\begin{equation}
\begin{split}
\square_{2d}=\frac{1}{4}\left[\frac{\partial}{\partial t^2}-ff'\frac{\partial}{\partial r}-f^2\frac{\partial}{\partial r^2}\right]+V(r),\\
V(r)=\frac{f}{4r^2}\left[l(l+1)+\frac{2M}{r}\right],
\end{split}
\end{equation}
the terms ${\cal M}^{(i)}_{\;(j)}$ are given
explicitly (omitting spherical harmonic indices $l,m$ unless needed for clarity) by
\begin{equation}
\begin{split}
    {\cal M}^{(1)}_{\;(j)}\bar h^{(j)}=&
\frac{1}{2}ff'\bar h^{(3)}_{,r}
+\frac{f}{2r^2}(1-4M/r)\left(\bar h^{(1)}-\bar h^{(5)}\right)\\
&
-\frac{1}{2r^2}\left[1-6M/r+12(M/r)^2\right]\bar h^{(3)}
+\frac{f^2}{2r^2}(6M/r-1)\bar h^{(6)},
\end{split}
\end{equation}
\begin{equation} \label{M2}
\begin{split}
{\cal M}^{(2)}_{\;(j)}\bar h^{(j)}=&
\frac{1}{2}ff'\bar h^{(3)}_{,r}
+\frac{f'}{2}\left(\bar h^{(2)}_{,t}-\bar h^{(1)}_{,t}\right)+\frac{ff'}{2}\left(\bar h^{(2)}_{,r}-\bar h^{(1)}_{,r}\right)
+\frac{f^2}{2r^2}\left(\bar h^{(2)}-\bar h^{(4)}\right)\\
&
+\frac{1}{2}(f'/r)\left[(1-4M/r)\bar h^{(3)}-f\left(\bar h^{(1)}-\bar h^{(5)}
-2f\bar h^{(6)}\right)\right],
\end{split}
\end{equation}
\begin{equation} \label{M3}
\begin{split}
  {\cal M}^{(3)}_{\;(j)}\bar h^{(j)}=&
\frac{1}{2}ff'\bar h^{(3)}_{,r}
+\frac{1}{2r^2}\left[1-8M/r+10(M/r)^2\right]\bar h^{(3)}\\
&
-\frac{f^2}{2r^2}\left[\bar h^{(1)}-\bar h^{(5)}-(1-4M/r)\bar h^{(6)}
\right],
\end{split}
\end{equation}
\begin{equation} \label{M4}
\begin{split}
{\cal M}^{(4)}_{\;(j)}\bar h^{(j)}=&
\frac{1}{4}f'\left(\bar h^{(4)}_{,t}-\bar h^{(5)}_{,t}\right)+\frac{ff'}{4}\left(\bar h^{(4)}_{,r}-\bar h^{(5)}_{,r}\right)
-\frac{1}{2}\,l(l+1)\,(f/r^2)\bar h^{(2)}\\
&
-\frac{1}{4}f'f/r\left[3\bar h^{(4)}+2\bar h^{(5)}-\bar h^{(7)}+l(l+1)\bar h^{(6)}\right],
\end{split}
\end{equation}
\begin{equation} \label{M5}
\begin{split}
{\cal M}^{(5)}_{\;(j)}\bar h^{(j)}=&
\frac{f}{r^2}\left[
(1-4.5M/r)\bar h^{(5)}-\frac{1}{2}l(l+1)\left(\bar h^{(1)}-\bar h^{(3)}\right)\right]\\
&+
\frac{f}{r^2}\left[\frac{1}{2}(1-3M/r)\left(l(l+1)\bar h^{(6)}-\bar h^{(7)}\right)
\right],
\end{split}
\end{equation}
\begin{equation} \label{M6}
{\cal M}^{(6)}_{\;(j)}\bar h^{(j)}=
-\frac{f}{2r^2}\left[\bar h^{(1)}-\bar h^{(5)}
-(1-4M/r)\left(f^{-1}\bar h^{(3)}+\bar h^{(6)}\right)\right],
\end{equation}
\begin{equation} \label{M7}
{\cal M}^{(7)}_{\;(j)}\bar h^{(j)}=
-\frac{f}{2r^2}\left(\bar h^{(7)}
+\lambda\,\bar h^{(5)}\right),
\end{equation}
\begin{equation} \label{M8}
{\cal M}^{(8)}_{\;(j)}\bar h^{(j)}=
\frac{f'}{4}\left(\bar h^{(8)}_{,t}-\bar h^{(9)}_{,t}\right)+\frac{ff'}{4}\left(\bar h^{(8)}_{,r}-\bar h^{(9)}_{,r}\right)
-\frac{1}{4}f'f/r\left(3\bar h^{(8)}+2\bar h^{(9)}-\bar h^{(10)}
\right),
\end{equation}
\begin{equation} \label{M9}
{\cal M}^{(9)}_{\;(j)}\bar h^{(j)}=
\frac{f}{r^2}\left(1-4.5M/r\right)\bar h^{(9)}
-\frac{f}{2r^2}\left(1-3M/r\right)\,\bar h^{(10)},
\end{equation}
\begin{equation} \label{M10}
{\cal M}^{(10)}_{\;(j)}\bar h^{(j)}=
-\frac{f}{2r^2}\left(\bar h^{(10)}+\lambda\,\bar h^{(9)}\right),
\end{equation}
$f'=2M/r^2$ and $\lambda=(l+2)(l-1)$.
The Lorenz gauge conditions read
\begin{equation}
L^1(t,r)=-\bar{h}_{,t}^{(1)}-\bar{h}_{,t}^{(3)}+f\left(\bar{h}_{,r}^{(2)}+\frac{\bar{h}^{(2)}-\bar{h}^{(4)}}{r}\right)=0,
\end{equation}
\begin{equation}
L^2(t,r)=\bar{h}_{,t}^{(2)}-f\bar{h}_{,r}^{(1)}+f\bar{h}_{,r}^{(3)}+(1-4/r)\bar{h}^{(3)}/r-\frac{f}{r}\left(\bar{h}^{(1)}-\bar{h}^{(5)}-f\bar{h}^{(3)}-2f\bar{h}^{(6)}\right)=0,
\end{equation}
\begin{equation}
L^3(t,r)=\bar{h}_{,t}^{(4)}-\frac{f}{r}\left(r\bar{h}_{,r}^{(5)}+2\bar{h}^{(5)}+l(l+1)\bar{h}^{(6)}-\bar{h}^{(7)}\right)=0,
\end{equation}
\begin{equation}
L^4(t,r)=\bar{h}_{,t}^{(8)}-\frac{f}{r}\left(r\bar{h}_{,r}^{(9)}+2\bar{h}^{(9)}-\bar{h}^{(10)}\right)=0.
\end{equation}
To calculate the decomposition of our singular field coordinate expansion into tensor
spherical-harmonic modes, we must evaluate the integrals of the singular field against the tensor spherical harmonics,
\begin{equation}
  \label{eq:hPilm-tr}
  \bar{h}^{(i)P}_{l m} = \frac{r}{\mu\, a^{(i)}_l} \int_{0}^{2 \pi} \int_{0}^{\pi} \bar{h}^P_{\tau\kappa} \tilde{\eta}^{\tau\mu}\tilde{\eta}^{\kappa\nu} Y_{\mu\nu}^{(i)l m}{}^\ast \sin \theta \, d\theta \, d\phi.
\end{equation}
The tensor spherical-harmonic modes of regular metric perturbation $\bar{h}^{R}_{\mu\nu}$ with the effective source are given by
\begin{equation}
\begin{split}
S^{(i)lm}_{\rm eff} &=\square_{2d}\overline{h}_{R}^{(i)lm}+\mathcal{M}^{(i)lm}_{\;(j)}\overline{h}_{R}^{(j)lm}\\
&=T^{(i)lm}\delta(r-r_p)-\square_{2d}\overline{h}_P^{(i)lm}-\mathcal{M}^{(i)lm}_{\;(j)}\overline{h}_P^{(j)lm}.
\end{split}
\end{equation}
Unlike the fundamental approach to mode decomposition for scalar-field treatment presented in Refs.~\cite{Warburton:2013lea,Heffernan:2017cad,Detweiler:2002mi,Barack:1999wf},
the gravitational field introduces several nontrivial complications \cite{Wardell:2015ada}.
Firstly, the use of tensor spherical harmonics rather than scalar harmonics significantly increases the complexity of the mode decomposition integrals.
Secondly, the implementation of the coordinate rotation technique that maps the particle's position to the North pole contributes only $m=0$ mode for the scalar field but various $m=0,1,2,3$ values for the gravitational field, reflecting the richer tensor structure.
Thirdly, the gravitational singular field develops the directional sensitivity of tensor components, absent in the scalar case due to the isotropy symmetry \cite{Wardell:2015ada}.
The first two items above add huge extra algebraic complexity to the problem and the third item introduces a spurious component in the puncture which makes the sum over modes converge very slowly.
Fortunately, there is a straightforward technique to these problem based on formulation in terms of a tetrad decomposition of the singular field \cite{Haas:2006ne}.
This method was originally developed for computing regularization parameters in mode-sum calculations, and its core methodology can be extended to the computation of effective sources in gravitational field calculations.

The necessary steps in computing the effective source are outlined as follows:
\begin{enumerate}
\item Write the singular field $\bar{h}^P_{\mu\nu}$ in $(t,r,\theta,\phi)$ coordinate components;
\item  Project the singular field $\bar{h}^P_{\mu\nu}$ into an orthonormal tetrad inertial frame;
\item  Modify the singular field components by different factors of $\cos\delta\theta$ in order to use the identities to simplify the angular integration;
\item  Rewrite the remaining integrals in terms of sums of $\psi^{lm}$ (where this was calculated by others previously for the scalar case) with coefficients that are functions of $r$ and the orbital parameters;
\item  Use the calculated coupling coefficients to project back into Barack-Lousto modes.
\end{enumerate}

We introduce an orthonormal tetrad frame $e^\alpha_{(\mu)}$ adapted to the spherical symmetry of the background spacetime.
At each point in the Schwarzschild geometry, this locally inertial frame consists of \cite{Haas:2006ne}
\begin{eqnarray}
e^\alpha_{(0)} &=& \biggl[ \frac{1}{\sqrt{f}}, 0, 0, 0 \biggr],  \\
e^\alpha_{(1)}=e^\alpha_{(+)} &=& \biggl[ 0, \sqrt{f}\sin\theta e^{+i\phi},
\frac{1}{r} \cos\theta e^{+i\phi},
\frac{+ie^{+i\phi}}{r\sin\theta} \biggr],  \\
e^\alpha_{(2)}=e^\alpha_{(-)} &=& \biggl[ 0, \sqrt{f}\sin\theta e^{-i\phi},
\frac{1}{r} \cos\theta e^{-i\phi},
\frac{-ie^{-i\phi}}{r\sin\theta} \biggr], \\
e^\alpha_{(3)} &=& \biggl[ 0, \sqrt{f}\cos\theta,
-\frac{1}{r} \sin\theta, 0 \biggr],
\end{eqnarray}
where Greek indices $\alpha,\beta,...$ denote spacetime components while parenthesized indices $(\mu)={(0),(1),(2),(3)}$ label the tetrad legs.
The corresponding dual tetrad $e_\alpha^{(\mu)}$ is defined by
\begin{eqnarray}
e_\alpha^{(0)} &=& \biggl[ f\frac{1}{\sqrt{f}}, 0, 0, 0 \biggr], \\
e_\alpha^{(1)}=e_\alpha^{(+)} &=& \frac12 \biggl[ 0, \frac{1}{f}\sqrt{f}\sin\theta e^{-i\phi},
r^2\frac{1}{r} \cos\theta e^{-i\phi},
r^2\sin^2\theta\frac{-ie^{-i\phi}}{r\sin\theta} \biggr],  \\
e_\alpha^{(2)}=e_\alpha^{(-)} &=& \frac12 \biggl[ 0, \frac{1}{f}\sqrt{f}\sin\theta e^{+i\phi},
r^2\frac{1}{r} \cos\theta e^{+i\phi},
r^2\sin^2\theta\frac{+ie^{+i\phi}}{r\sin\theta} \biggr],  \\
e_\alpha^{(3)} &=& \biggl[ 0, \frac{1}{f}\sqrt{f}\cos\theta,
-r^2\frac{1}{r} \sin\theta, 0 \biggr].
\end{eqnarray}
The tetrad-projected components
\begin{equation}
\bar{h}^P_{(\mu)(\nu)}=e^\alpha_{(\mu)}e^\beta_{(\nu)} \bar{h}^P_{\alpha\beta}
\end{equation}
transform the tensor perturbation \(\bar{h}^P_{\alpha\beta}\) into a set of scalar functions that are particularly well-suited for analysis in spherically symmetric spacetimes. These projected components inherit their scalar character from the orthonormal tetrad frame $e_{(\mu)}^\alpha$, with each $\bar{h}^P_{(\mu)(\nu)}$ behaving as a genuine scalar field under coordinate transformations.

This scalar property enables a natural decomposition in terms of scalar spherical harmonics $Y_{l m}(\theta,\phi)$, which provides several key advantages.
Firstly, the harmonic expansion completely decouples the angular dependence from the radial-temporal evolution of the perturbation.
Secondly, the decomposition automatically handles the regularity conditions at the poles that would otherwise complicate tensor harmonic treatments.
The tetrad projection thus serves as a crucial bridge between the geometric, coordinate-independent nature of the tensor perturbation and the practical requirements for numerical computation and regularization in self-force calculations.
This approach preserves all physical information while avoiding the coordinate singularities and gauge complications that arise in direct tensor treatments of the perturbation field.

The spherical-harmonic modes of each tetrad frame component can be written (omitting the indices $P$ for brevity),
\begin{equation}\label{eq:hmunulm}
  \bar{h}_{(\mu)(\nu)}=\sum_{lm}  \bar{h}^{lm}_{(\mu)(\nu)}(t,r)Y_{lm}(\theta,\phi).
\end{equation}
In order to determine the relation between $\bar{h}^{lm}_{(\mu)(\nu)}$, the modes of the frame components, and $\bar{h}^{(i)}_{lm}$, the modes of the original tensor field.
Substituting Eq.~\eqref{eq:hmunulm} into Eq.~\eqref{eq:hPilm-tr} yields
\begin{equation}\label{eq:cij}
\begin{split}
  \bar{h}^{(i)}_{lm}&=\frac{r}{a^{(i)}_l}\bar{h}^{l'm'}_{(\mu)(\nu)}\int Y_{l'm'}(\theta,\phi)e^{(\mu)}_{\alpha}e^{(\nu)}_{\beta}\tilde{\eta}^{\alpha\gamma}\tilde{\eta}^{\beta\xi}\bar{Y}_{\gamma\xi}^{(i)lm}(\theta,\varphi;r)  d\Omega,\\
  &=\frac{r}{a^{(i)}_l}C^{(i)(\mu)(\nu)}_{\left( lm|l'm' \right)}\, \bar{h}^{l'm'}_{(\mu)(\nu)}\left[(\theta,\phi)~\rm{decomposition}\right].
\end{split}
\end{equation}
The coupling coefficients $C^{(i)(\mu)(\nu)}_{\left( lm|l'm' \right)}$ are given analytically in Appendix.~\ref{AppB}.
In what follows, we present the systematic method to  derive the spherical harmonic decomposition $\bar{h}^{lm}_{(\mu)(\nu)}$ for the tetrad components $\bar{h}_{(\mu)(\nu)}$.
Beginning with the component $\bar{h}_{(0)(0)}$, its harmonic expansion takes the form
\begin{equation}
\begin{split}
 \bar{h}^{lm}_{(0)(0)}&=\int \bar{h}_{(0)(0)}  Y^{lm\ast} (\theta,\phi)d\Omega\\
 &=\int \left(\frac{u_tu_t}{f}+\frac{2u_tu_t}{(r_p-2)r_p f} \right)\frac{1}{\bar{s}} Y^{lm\ast} (\theta,\phi)d\Omega\\
 &=\left(\frac{u_tu_t}{f}+\frac{2u_tu_t}{(r_p-2)r_p f} \right)\int \frac{1}{\bar{s}} Y^{lm\ast} (\theta,\phi)d\Omega.
\end{split}
\end{equation}
The analysis reveals that the integral expression,
\begin{equation}
\psi^{lm}=r\int \frac{1}{\bar{s}} Y^{lm\ast} (\theta,\phi)d\Omega,
\end{equation}
precisely corresponds to the harmonic decomposition of a scalar singular field.
This observation naturally leads us to investigate whether all tetrad components $\bar{h}^{lm}_{(\mu)(\nu)}$
can be similarly expressed through such scalar harmonic projections.
Building upon previous extensive analysis of the scalar singular field, the harmonic decomposition $\psi^{lm}$ has been explicitly derived for generic orbits in Schwarzschild spacetime \cite{Heffernan:2017cad,Detweiler:2002gi,Leather:2023dzj,Heffernan:2017cad}.
We will prove that such a representation is indeed possible for all tetrad components $\bar{h}^{lm}_{(\mu)(\nu)}$.
Let us examine another component, $\bar{h}_{(0)(1)}$, whose harmonic decomposition takes the form
\begin{equation}
 \bar{h}^{lm}_{(0)(1)}=\int \bar{h}_{(0)(1)}  Y^{lm\ast} (\theta,\phi)d\Omega,
\end{equation}
where the explicit integrand $\bar{h}_{(0)(1)}$ is expressed as
\begin{equation}
\begin{split}
\bar{h}_{(0)(1)}=&n_{01,1}\frac{\csc\theta }{\bar{s}}
+n_{01,2}\frac{\sin\theta }{\bar{s}}
+n_{01,3}\frac{e^{2i\phi}\csc\theta }{\bar{s}}\\
&
+n_{01,4}\frac{e^{2i\phi}}{\bar{s}}
+n_{01,5}\frac{e^{i\phi}\sin\theta}{\bar{s}}
+n_{01,6}\frac{e^{i\phi}\csc\theta}{\bar{s}}+n_{01,7}\frac{e^{i\phi}\cos^2\theta}{\bar{s}},
\end{split}
\end{equation}
where the coefficients $n_{01,1}\sim n_{01,7}$ are independent of angular coordinates $\theta$ and $\phi$.
It seems that $\bar{h}^{lm}_{(0)(1)}$ cannot be directly expressed in terms of the scalar singular field harmonics $\psi^{lm}$.
This limitation becomes apparent when comparing its harmonic projection with the idealized form that would allow decomposition via
\begin{equation}\label{h01}
\bar{h}^{lm}_{(0)(1)} \stackrel{?}{=} \int \sum_{ij}\frac{a_{ij}(r)}{\bar{s}} Y^{l+i,m+j \ast} (\theta,\phi)d\Omega =\sum_{ij} a_{ij}(r)\psi^{l+i,m+j}.
\end{equation}
The obstruction arises because the angular structure couples differently with the radial dependence.
To achieve the desired harmonic decomposition in Eq.~\eqref{h01} while maintaining second-order distance accuracy in the local expansion, we systematically modify the puncture field components through angular modulation,
\begin{equation}
1\to \cos\delta\theta~[=1+\mathcal{O}(\delta\theta^2)].
\end{equation}
For example, the transformed  component $\chi_{tr}$ becomes
\begin{equation}
\chi_{tr}= u_tu_r  \cos\delta\theta\cos\delta\theta+\frac{u_tu_\phi }{r_p}w^3 \cos\delta\theta.
\end{equation}
This adjustment strategy extends to all components through the procedure by multiplying by multiple $\cos\delta\theta$ factors in Appendix.~\ref{AppC}.
Through this formulation, we obtain the transformed
$\bar{h}_{(0)(1)}$ in the form
\begin{equation}
\bar{h}_{(0)(1)}=s_{01,1}\frac{\sin ^2\theta }{\bar{s}}+s_{01,2}
\frac{\sin ^2\theta e^{2 i \phi}}{\bar{s}}+s_{01,3}
\frac{\sin ^3\theta  e^{i \phi}}{\bar{s}}+s_{01,4}
\frac{\sin\theta  e^{ i \phi}}{\bar{s}}+s_{01,5}
\frac{\cos^2\theta\sin\theta e^{i \phi}}{\bar{s}},
\end{equation}
where the coefficients are
\begin{equation}
\begin{split}
s_{01,1}&= \frac{(r_p-2) r_p u_r +i\,r\,u_\phi  \sqrt{f}}{2 r r_p\sqrt{f}}u_t e^{i \phi_p},\\
s_{01,2}&= \frac{ (r_p-2) r_p u_r +i r u_\phi  \sqrt{f}}{2 r r_p\sqrt{f}}u_t  e^{-i \phi_p},\\
s_{01,3}&= u_t  u_r  ,\\
s_{01,4}&= i\frac{  r_p^2+r_p (\delta r-2)-\delta r}{r (r_p-2) r_p \sqrt{f}}u_t u_\phi ,\\
s_{01,5}&= \frac{r_p-2 }{r  \sqrt{f}}u_t u_r  .
\end{split}
\end{equation}
By applying the standard spherical harmonic identities
\begin{equation}
\begin{split}
\cos\theta  Y^{lm\ast} =&
\sqrt{ \frac{(l-m+1)(l+m+1)}{(2l+1)(2l+3)} } Y^{l+1,m\ast}
+ \sqrt{ \frac{(l-m)(l+m)}{(2l-1)(2l+1)} } Y^{l-1,m\ast},  \\
\sin\theta e^{-i\phi} Y^{lm\ast} =&
-\sqrt{ \frac{(l+m+1)(l+m+2)}{(2l+1)(2l+3)} } Y^{l+1,m+1\ast}
+ \sqrt{ \frac{(l-m)(l-m-1)}{(2l-1)(2l+1)} } Y^{l-1,m+1\ast},  \\
\sin\theta e^{+i\phi} Y^{lm\ast} =&
\sqrt{ \frac{(l-m+1)(l-m+2)}{(2l+1)(2l+3)} } Y^{l+1,m-1\ast}
- \sqrt{ \frac{(l+m)(l+m-1)}{(2l-1)(2l+1)} } Y^{l-1,m-1\ast},
\end{split}\label{standardidentity}
\end{equation}
the transformed $\bar{h}_{(0)(1)}$ can indeed be expressed through scalar singular field harmonics
\begin{equation}\label{eq:aij}
\begin{split}
\bar{h}^{lm}_{(0)(1)} & =\int \bar{h}_{(0)(1)}  Y^{lm\ast} (\theta,\phi)d\Omega\\
&=\int \sum_{ij} \frac{a_{ij}(r)}{\bar{s}} Y^{l+i,m+j*} (\theta,\phi) d\Omega =\sum_{ij} a_{ij}(r) \psi^{l+i,m+j}.
\end{split}
\end{equation}
This key result can extend to all tetrad components $\bar{h}^{lm}_{(\mu)(\nu)}$, establishing that: 1)
The complete gravitational singular metric field can be reconstructed from scalar harmonics;
2) Different modes-coupling is systematically controlled through the coefficients $a_{ij}(r)$;
3) The harmonic decomposition maintains the singular field's local behavior.
So the computational procedure should firstly compute the scalar harmonic modes $\psi^{lm}$ and then apply the derived analytical transformation coefficients $a_{ij}(r)$ in Eq.~\eqref{eq:aij} together with the coupling coefficients $C^{(i)(\mu)(\nu)}_{\left( lm|l'm' \right)}$ in Eq.~\eqref{eq:cij} to construct all ten components $h^{(i)lm}$ through linear combination.
As a concrete example, consider the component $h^{(1)lm}$ for the specific case of $l=2$ and $m=2$. Its analytical expression takes the form
\begin{equation}
h^{(1)l=2,m=2}= d_{11} \,\psi^{l=1,m=1} + d_{22}\, \psi^{l=2,m=2} + d_{31}\, \psi^{l=3,m=1} + d_{33}\, \psi^{l=3,m=3},
\end{equation}
where the coefficients are given by
\begin{equation}
\begin{split}
d_{11} &= 14 i \sqrt{5} (r-2)^2 (r_p-2)^2 u^r L e^{-i \phi_p}, \\
d_{22} &= \frac{4 r_p^3 (u^r)^2}{r^2 (r_p-2)^3} - \frac{4 r_p^3 (u^r)^2}{r (r_p-2)^3} - \frac{2 r r_p (u^r)^2}{(r_p-2)^3} - \frac{8 r_p (u^r)^2}{r (r_p-2)^3} \\
&\quad + \frac{2 r r_p E^2}{(r_p-2)^3} + \frac{8 r E^2}{(r_p-2)^3 r_p} - \frac{8 r E^2}{(r_p-2)^3} + \frac{r_p^3 (u^r)^2}{(r_p-2)^3} \\
&\quad + \frac{r_p^3 E^2}{(r_p-2)^3} - \frac{8 r_p^2 E^2}{(r_p-2)^3} + \frac{8 r_p (u^r)^2}{(r_p-2)^3} + \frac{20 r_p E^2}{(r_p-2)^3} - \frac{16 E^2}{(r_p-2)^3}, \\
d_{31} &= -i \sqrt{70} (r-2)^2 (r_p-2)^2 u^r L e^{-i \phi_p}, \\
d_{33} &= -5 i \sqrt{42} (r-2)^2 (r_p-2)^2 u^r L e^{i \phi_p},
\end{split}
\end{equation}
with $E$ and $L$ representing the particle's energy and angular momentum, respectively.
For brevity, we have not included the complete set of mathematical expressions here, but all relevant formulas are available in \url{https://github.com/ChaoZhangCode/EffectiveSourceCode}.
The harmonic components of the  scalar field $\psi^{lm}$ have already been given in Refs.~ \cite{Heffernan:2017cad,Detweiler:2002gi,Leather:2023dzj,Heffernan:2017cad}.
For our analysis, we employ the puncture scheme  developed in Ref.~\cite{Heffernan:2017cad} and the full expression is given explicitly in the supplemental material of paper \cite{Leather:2023dzj} for a Mathematica notebook containing the high-order distance puncture.
The complete analytic expression for the effective source is designed for application in both frequency-domain and time-domain calculations.
For frequency-domain applications, we anticipate no significant challenges for generic orbits.
The procedure developed in standard mode-sum method  can be applied to obtain the homogeneous solution \cite{Akcay:2013wfa}, and the well-established extended source method may then be used to compute the corresponding inhomogeneous solution and the resulting GSF \cite{Leather:2023dzj}.
In principle, this method of deriving effective source can be extended to higher-order punctures in the distance parameter, introducing higher powers of $\bar{s}$ to further smooth the effective source and accelerate convergence of the mode.
Such an extension first requires the derivation of higher-order punctures for the scalar field.
The tetrad components must then be substituted up to the corresponding order to align their form with the scalar results.
This procedure, though highly challenging, ultimately produces higher-order punctures that achieve improved accuracy in the distance expansion.

\section{Effective Source}\label{sec6}
\begin{figure*}[htp]
  \centering
  \includegraphics[width=0.9\textwidth]{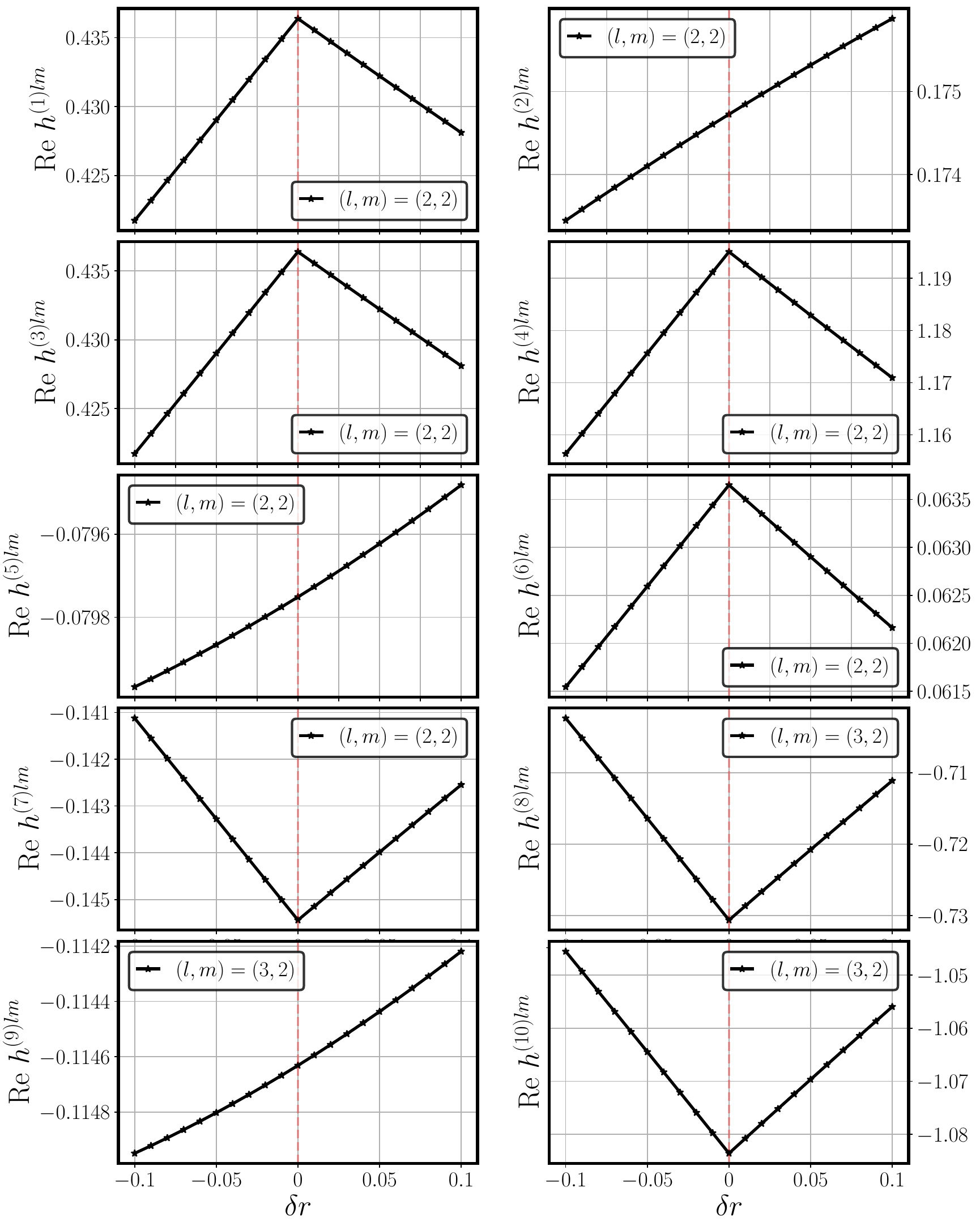}
  \caption{Circular case:
Real components of the puncture field modes $\bar{h}^{(i)lm}$ for a particle on a circular orbit at $r_p=10~M$ and $\phi_p=\pi/6$. Top panels display even-parity components $\bar{h}^{(1-7)}$ for $(l,m)=(2,2)$, showing continuous variation across the particle's position (vertical dashed line at $\delta r=0$). Bottom panels show odd-parity components $\bar{h}^{(8-10)}$ for $(l,m)=(3,2)$, exhibiting characteristic parity-dependent behavior.
  }
  \label{fig:hiC}
\end{figure*}

\begin{figure*}[htp]
  \centering
  \includegraphics[width=0.9\textwidth]{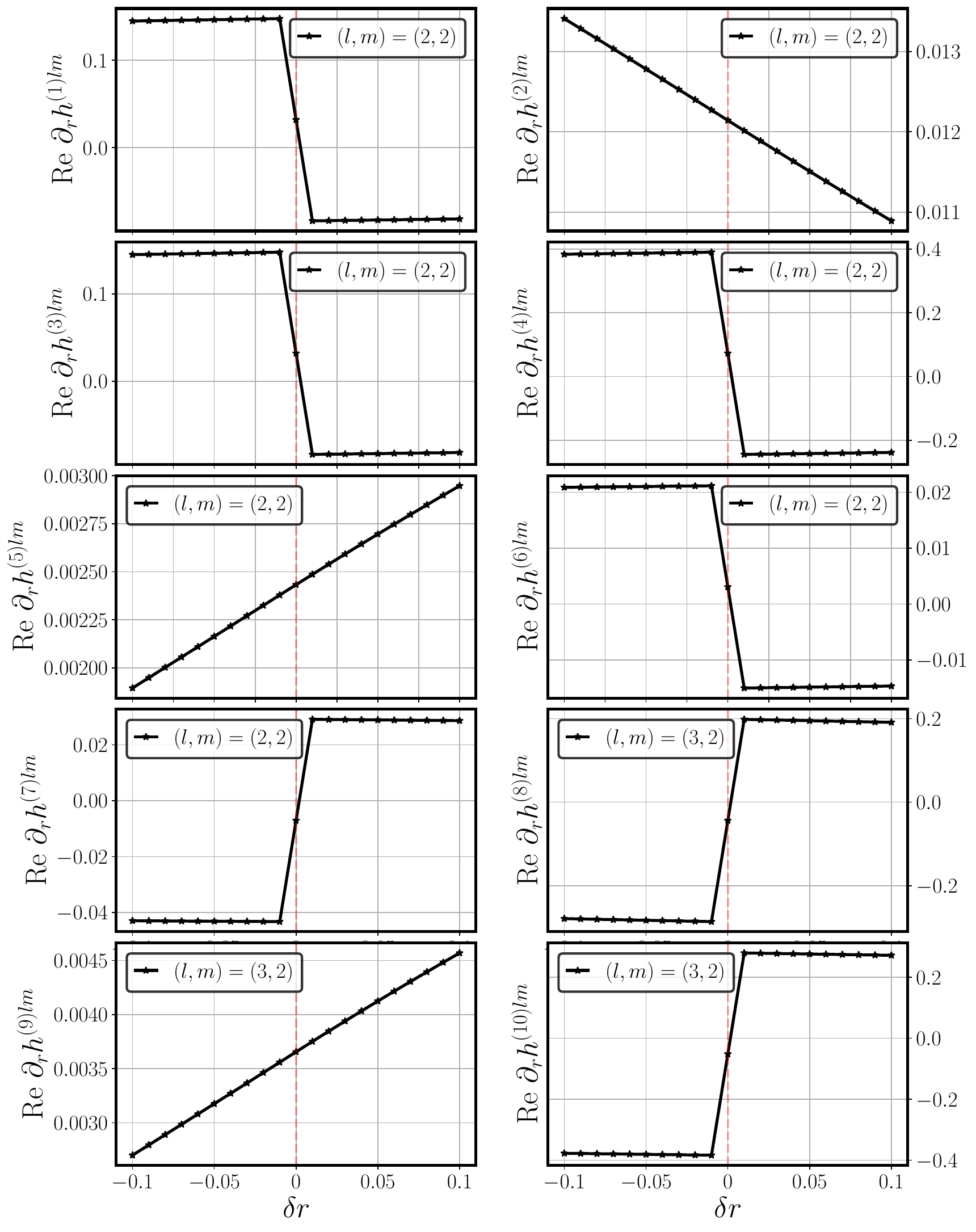}
  \caption{Circular case:
  Real components of the puncture field with respect to $r$ modes $\partial_r \bar{h}^{(i)lm}$  for a particle on a circular orbit at $r_p=10~M$ and $\phi_p=\pi/6$. Top panels display even-parity components $\bar{h}^{(1-7)}$ for $(l,m)=(2,2)$ and the bottom panels show odd-parity components $\bar{h}^{(8-10)}$ for $(l,m)=(3,2)$, exhibiting characteristic parity-dependent behavior.
  }
  \label{fig:dhidrC}
\end{figure*}

\begin{figure*}[htp]
  \centering
  \includegraphics[width=0.9\textwidth]{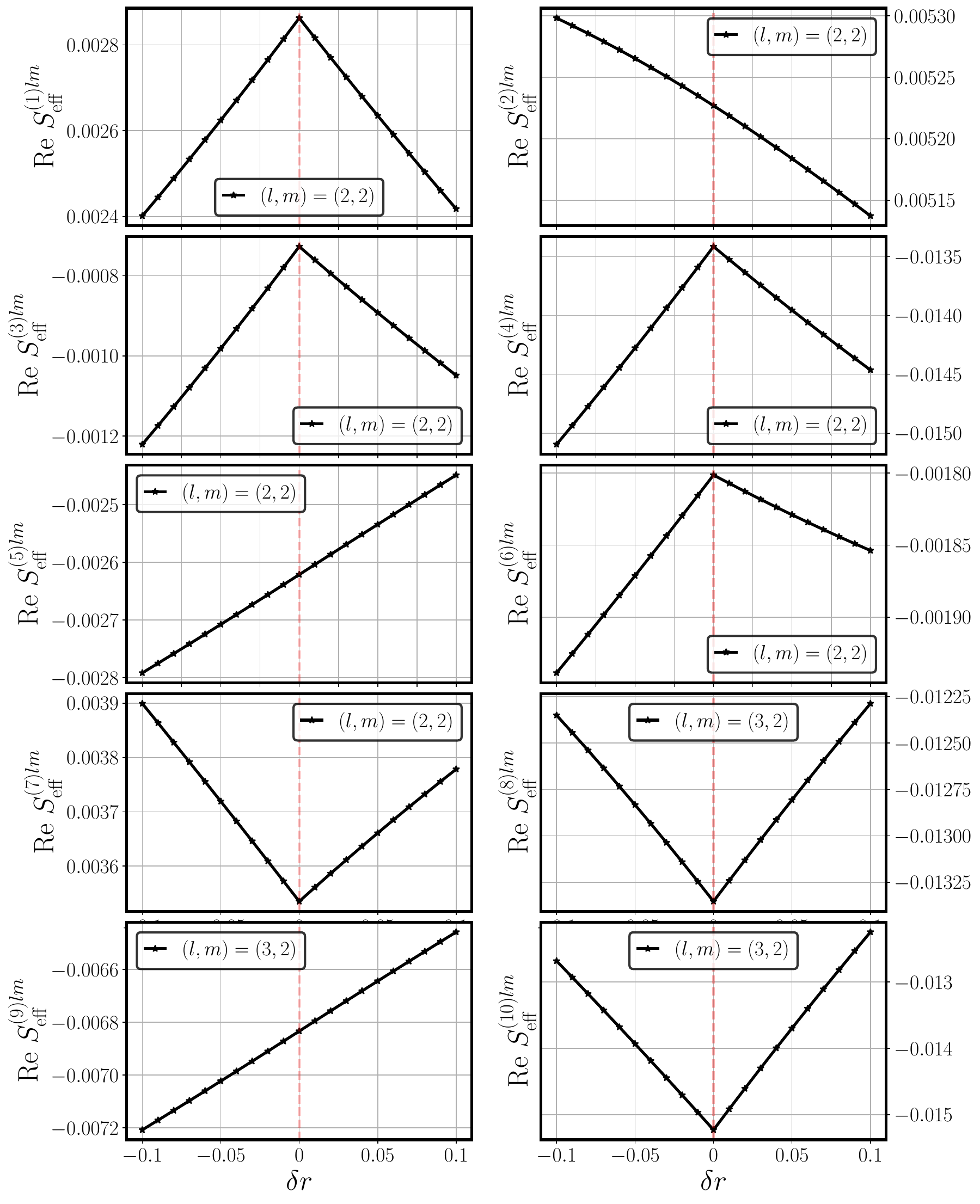}
  \caption{Circular case:
Real components of the effective source modes $S_{\rm eff}^{(i)lm}$ for a particle on a circular orbit at $r_p=10~M$ and $\phi_p=\pi/6$. Top panels display even-parity components $\bar{h}^{(1-7)}$ for $(l,m)=(2,2)$, showing continuous variation across the particle's position (vertical dashed line at $\delta r=0$). Bottom panels show odd-parity components $\bar{h}^{(8-10)}$ for $(l,m)=(3,2)$, exhibiting characteristic parity-dependent behavior.
All components remain continuous and finite at $\delta r=0$, validating the regularization scheme.
  }
  \label{fig:efsourceC}
\end{figure*}

With an approximation to the gravitational singular field at hand, we can now calculate the corresponding effective source.
The effective source near the particle position are given by
\begin{equation}
S^{(i)lm}_{\rm eff}=T^{(i)lm}\delta(r-r_p)-\square_{2d}\overline{h}_P^{(i)lm}-\mathcal{M}^{(i)l}_{\;(j)}\overline{h}_P^{(j)lm}.
\end{equation}
In Figs. \ref{fig:hiC} and \ref{fig:dhidrC}, the numerical results are presented for the gravitational singular field tensor harmonics $\bar{h}^{(i)lm}$ and the corresponding harmonics with respect to $r$ for a circular orbit at the position $r_p=10~M$ and $\phi_p=\pi/6$.
The field components naturally separate into two distinct parity classes that components $\bar{h}^{(i=1,...,7)lm}$ possess even parity (invariant under spatial inversion), while $\bar{h}^{(i=8,9,10)lm}$ exhibit odd parity (sign-changing under inversion).
This parity distinction has important consequences for their mode structure; even-parity components vanish for odd $l+m$, while odd-parity components vanish for even $l+m$.
Figure \ref{fig:hiC} displays the field components near the particle's position, with $\bar{h}^{(1-7)lm}$ shown for $(l,m)=(2,2)$ and $\bar{h}^{(8-10)lm}$ for $(l,m)=(3,2)$.
The vertical dashed line marks the particle's location at $r_p=10~M$.
Figure \ref{fig:dhidrC} presents the radial derivatives of these components, demonstrating the expected nondifferentiability at the particle position for $\bar{h}^{(1,3,4,6,7,8,10)lm}$ while maintaining continuity for $\bar{h}^{(2,5,9)lm}$- behavior consistent with previous circular-orbit studies \cite{Wardell:2015ada}.
The effective source, plotted in Fig. \ref{fig:efsourceC}, exhibits the crucial properties required for successful self-force calculations: it remains finite and continuous (though not necessarily differentiable) across the particle's worldline.

\begin{figure*}[htp]
  \centering
  \includegraphics[width=0.9\textwidth]{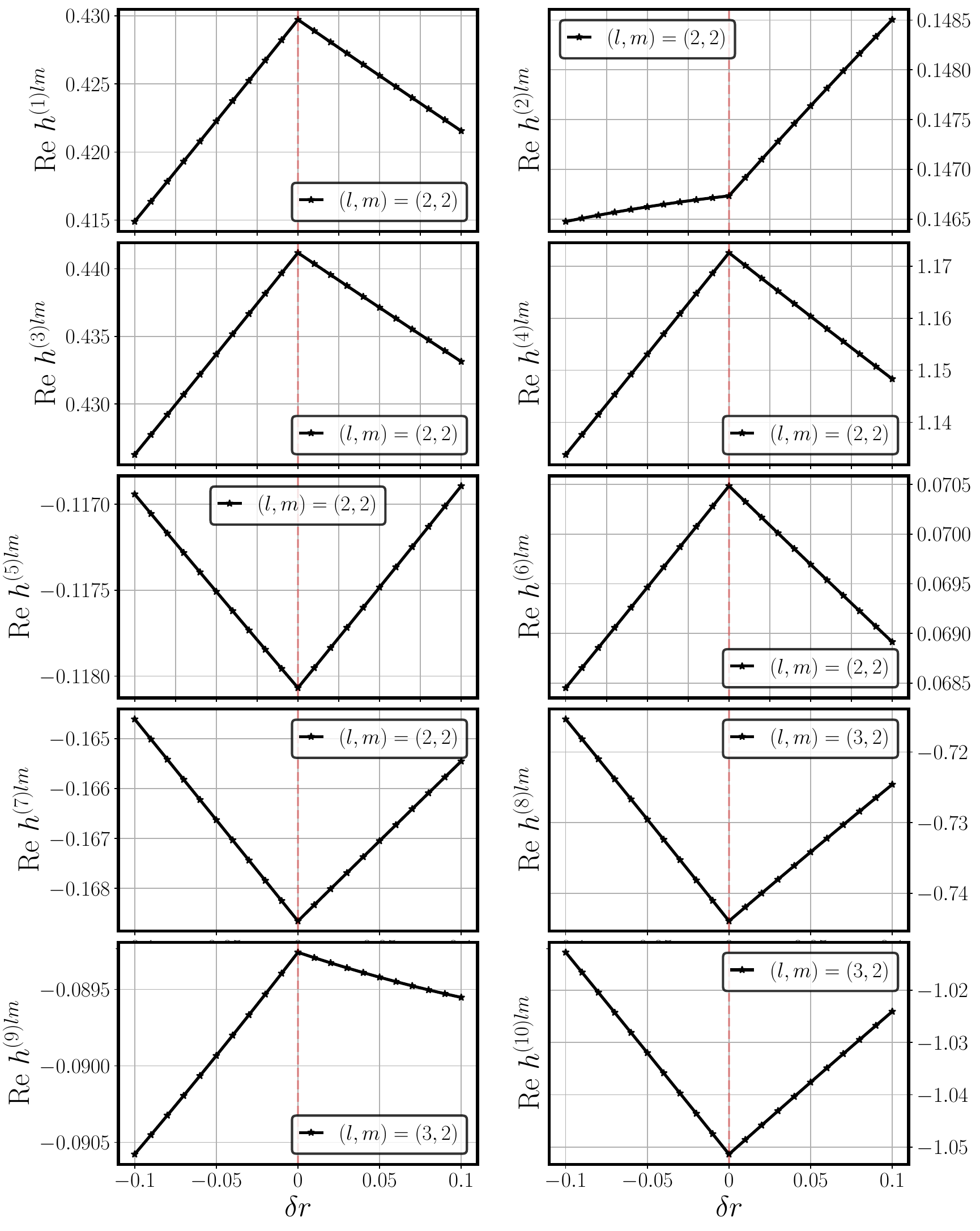}
  \caption{Eccentric case:
Real components of the puncture field modes $\bar{h}^{(i)lm}$ for a particle on an eccentric orbit with the velocity $u^r=\frac{1}{3 \sqrt{95}}$ and $u^\phi=\frac{7}{30 \sqrt{38}}$ at $r_p=10~M$ and $\phi_p=\pi/6$. Top panels display even-parity components $\bar{h}^{(1-7)}$ for $(l,m)=(2,2)$, showing continuous variation across the particle's position (vertical dashed line at $\delta r=0$). Bottom panels show odd-parity components $\bar{h}^{(8-10)}$ for $(l,m)=(3,2)$, exhibiting characteristic parity-dependent behavior.
  }
  \label{fig:hi}
\end{figure*}

\begin{figure*}[htp]
  \centering
  \includegraphics[width=0.9\textwidth]{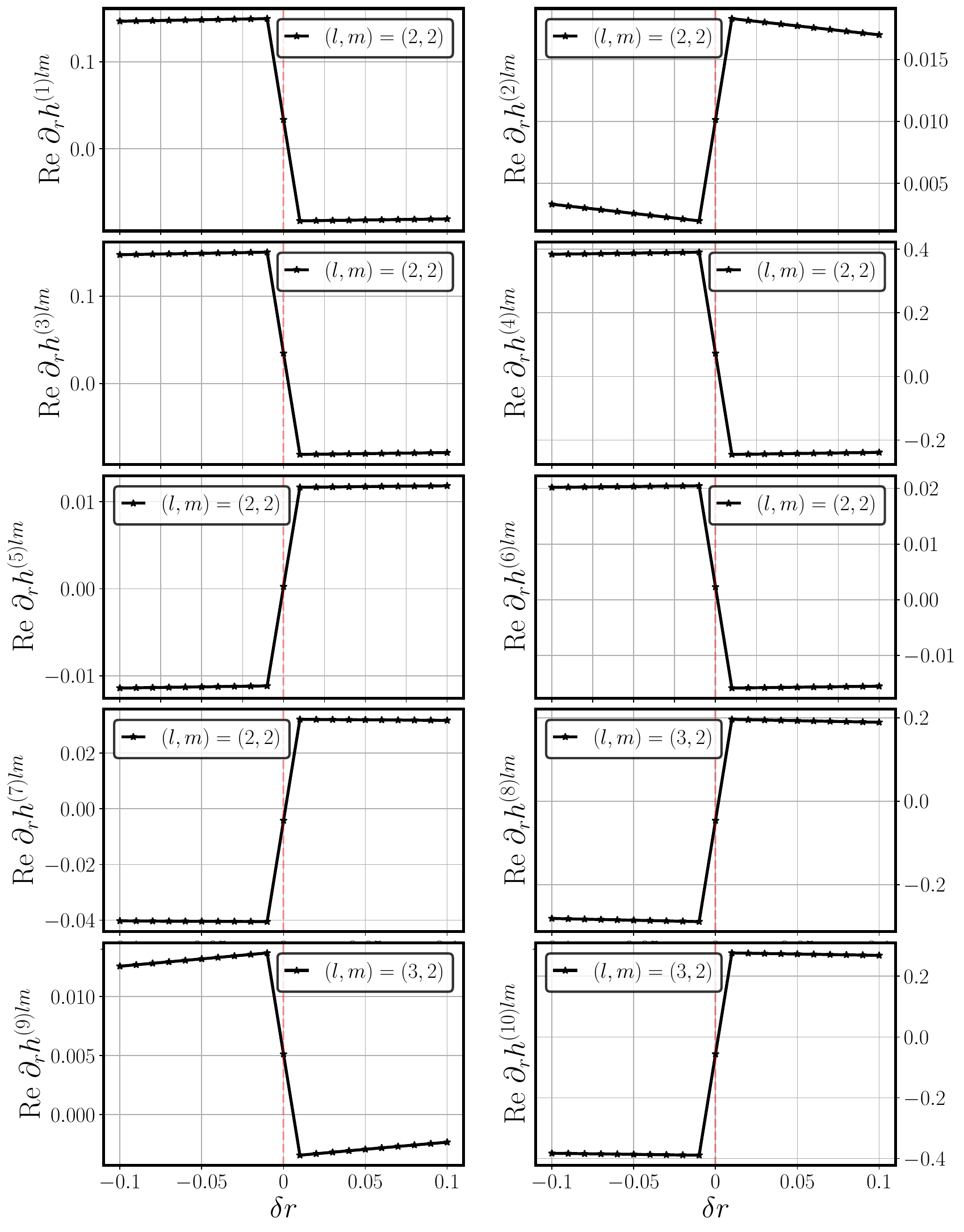}
  \caption{Eccentric case:
  Real components of the puncture field with respect to $r$ modes $\partial_r \bar{h}^{(i)lm}$  for a particle on an eccentric orbit with the velocity $u^r=\frac{1}{3 \sqrt{95}}$ and $u^\phi=\frac{7}{30 \sqrt{38}}$ at $r_p=10~M$ and $\phi_p=\pi/6$. Top panels display even-parity components $\bar{h}^{(1-7)}$ for $(l,m)=(2,2)$ and the bottom panels show odd-parity components $\bar{h}^{(8-10)}$ for $(l,m)=(3,2)$.
  All components remain nondifferentiability at $\delta r=0$, validating the regularization scheme.}
  \label{fig:dhidr}
\end{figure*}

\begin{figure*}[htp]
  \centering
  \includegraphics[width=0.9\textwidth]{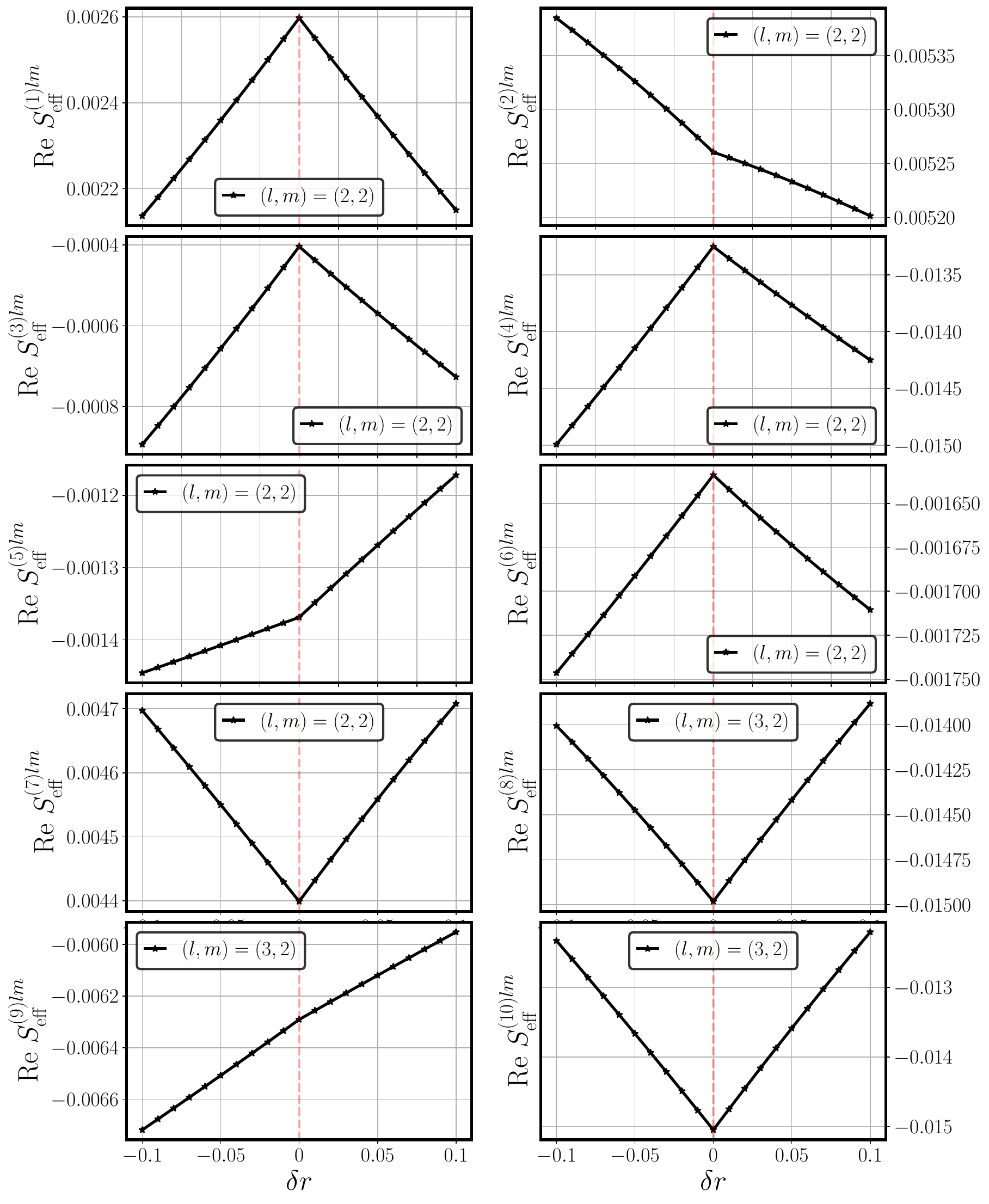}
  \caption{Eccentric case:
Real components of the effective source modes $S_{\rm eff}^{(i)lm}$ for a particle on an eccentric orbit with the velocity $u^r=\frac{1}{3 \sqrt{95}}$ and $u^\phi=\frac{7}{30 \sqrt{38}}$ at $r_p=10~M$ and $\phi_p=\pi/6$. Top panels display even-parity components $\bar{h}^{(1-7)}$ for $(l,m)=(2,2)$, showing continuous variation across the particle's position (vertical dashed line at $\delta r=0$). Bottom panels show odd-parity components $\bar{h}^{(8-10)}$ for $(l,m)=(3,2)$, exhibiting characteristic parity-dependent behavior.
All components remain finite at $\delta r=0$, validating the regularization scheme.
  }
  \label{fig:efsource}
\end{figure*}

\begin{figure*}[htp]
  \centering
  \includegraphics[width=0.9\textwidth]{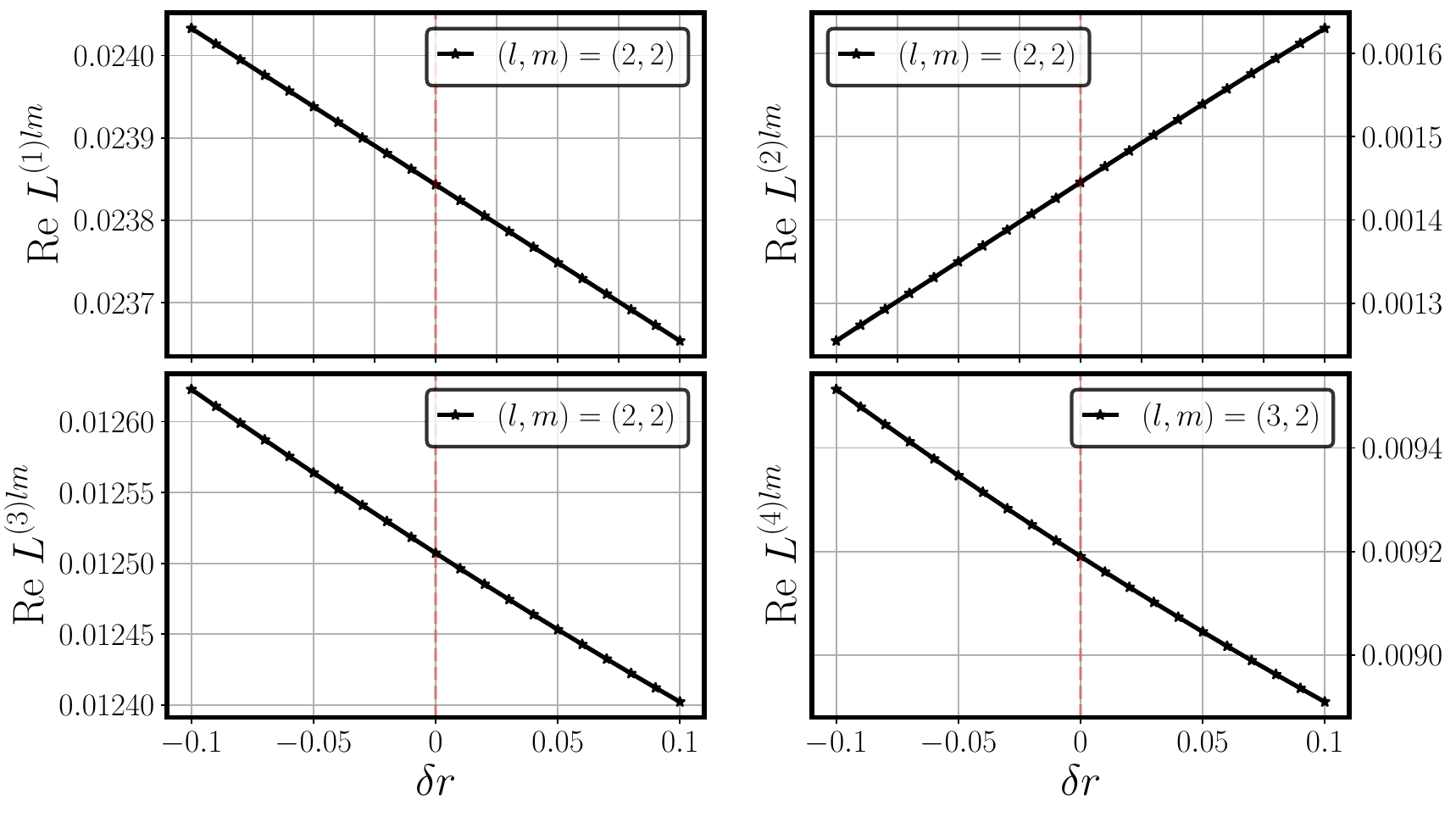}
  \caption{Eccentric case:
Real components of the Lorenz condition for a particle on an eccentric orbit with the velocity $u^r=\frac{1}{3 \sqrt{95}}$ and $u^\phi=\frac{7}{30 \sqrt{38}}$ at $r_p=10~M$ and $\phi_p=\pi/6$.
The results show $C^1$ continuous variation across the particle's position (vertical dashed line at $\delta r=0$), validating the regularization scheme.
  }
  \label{fig:lorenz}
\end{figure*}

For an eccentric orbit at the position $r_p=10~M$ and $\phi_p=\pi/6$ with the velocity $u^r=\frac{1}{3 \sqrt{95}}$ and $u^\phi=\frac{7}{30 \sqrt{38}}$, the numerical results for the gravitational singular field harmonics $\bar{h}^{(i)lm}$ and the gravitational singular field with respect to $r$ are presented in Figs. \ref{fig:hi} and \ref{fig:dhidr}.
There is a similar result with the circular case,
the field components can separate into even parity for $\bar{h}^{(i=1,...,7)lm}$ and odd parity for $\bar{h}^{(i=8,9,10)lm}$.
Figure \ref{fig:hi} displays the puncture field components near the particle's position, with $\bar{h}^{(1-7)lm}$ shown for $(l,m)=(2,2)$ and $\bar{h}^{(8-10)lm}$ for $(l,m)=(3,2)$.
The vertical dashed line marks the particle's location at $r_p=10~M$.
Contrast with the circular case, the radial derivatives of field have the nondifferentiability at the particle position for all components $\bar{h}^{(1-10)lm}$ that $\bar{h}^{(2,5,9)lm}$ will not maintain continuity at the particle position for the eccentric orbit in Fig.~\ref{fig:dhidr}.
The effective source, plotted in Fig. \ref{fig:efsource}, exhibits the same crucial properties required for successful self-force calculations; it remains finite and continuous (though not necessarily differentiable) across the particle's worldline.
Additionally, as plotted in Fig. \ref{fig:lorenz}, the Lorenz gauge condition exhibits $C^1$ continuity across the particle's worldline, which validates our regularization scheme.
This confirms that our method properly handles the challenges of eccentric orbits and provides a robust foundation for numerical solutions of the perturbed Einstein equations.
The results validate our implementation of the effective source approach for generic orbital configurations in black hole perturbation theory.


\section{Conclusion}\label{sec7}
In this paper, we establish a comprehensive framework for constructing a generic, finite, and continuous effective source for GSF calculations in Schwarzschild spacetime, overcoming the fundamental challenges posed by point-particle singularities in EMRIs.
By developing a novel analytical formulation that combines covariant singular field expansions with a tetrad-based spherical harmonic decomposition, we enable precise computation of effective source for arbitrary geodesic motion, including eccentric orbits.
The method's numerical validation demonstrates its robustness in maintaining finite and continuous behavior across the particle's worldline while significantly improving computational efficiency through optimized 1+1D implementations.
These advances provide a crucial foundation for future precision gravitational waveform modeling, particularly for space-based observations and strong-field tests of general relativity, while opening new avenues for extending this framework to self-consistent and second-order mass-ratio GSF calculations.

\begin{acknowledgments}
This work is supported in part by the National Key Research and Development Program of China under Grant No. 2020YFC2201504, 
and the National Natural Science Foundation of China key project under Grant No. 12535002.

\end{acknowledgments}

\appendix

\section{The Coupling Coefficients} \label{AppB}
The coupling coefficients $C^{(i)(\mu)(\nu)}_{\left( lm|l'm' \right)}$ are given as
\begin{equation}
\begin{split}
C^{(1)(0)(0)}_{\left( lm|l'm' \right)}&=\frac{\left(1-\frac{2}{r}\right)}{\sqrt{2}}\delta^{l}_{l'}\delta^{m}_{m'},\\
C^{(1)(0)(1)}_{\left( lm|l'm' \right)}&=C^{(1)(0)(2)}_{\left( lm|l'm' \right)}=C^{(1)(0)(3)}_{\left( lm|l'm' \right)}=C^{(1)(1)(0)}_{\left( lm|l'm' \right)}=0,\\
C^{(1)(1)(1)}_{\left( lm|l'm' \right)}&=\frac{f}{4 \sqrt{2}}\Big(\sqrt{\frac{(l-m-3) (l-m-2) (l-m-1) (l-m)}{(1-2 l)^2 (2 l-3) (2 l+1)}}\delta^{l-2}_{l'}\delta^{m+2}_{m'}\\
&+\sqrt{\frac{(l+m+1) (l+m+2) (l+m+3) (l+m+4)}{(2 l+1) (2 l+3)^2 (2 l+5)}}\delta^{l+2}_{l'}\delta^{m+2}_{m'} \\
&-\frac{2 \sqrt{(l-m-1) (l-m) (l+m+1) (l+m+2)}}{4 l (l+1)-3}\delta^{l}_{l'}\delta^{m+2}_{m'} \Big),\\
&\cdots
\end{split}
\end{equation}
where the complete set of mathematical expressions is not fully listed here due to space constraints, but all relevant formulas can be found in \url{https://github.com/ChaoZhangCode/EffectiveSourceCode}.

\section{The Transformed  Singular Field} \label{AppC}
The nonvanishing transformed components of $\chi_{\mu\nu}$ are
\begin{equation}
\begin{split}
\chi_{tt}&=u_tu_t+\frac{2u_tu_t}{r_p(r_p-2)}w^1,\\
\chi_{tr}&=u_tu_r \cos^2\delta\theta+\frac{u_tu_\phi }{r_p}w^3\cos\delta\theta,\\
\chi_{t\theta}&=-(r_p-2)u_tu_r w^2\cos\delta\theta,\\
\chi_{t\phi}&=-(r_p-2)u_tu_r w^3\cos^3\delta\theta+u_tu_\phi \cos^2\delta\theta+\frac{(r_p-1)u_tu_\phi }{(r_p-2)r_p}w^1\cos^2\delta\theta,\\
\chi_{rr}&=u_r u_r -\frac{2u_r u_r }{(r_p-2)r_p}w^1+\frac{2u_r u_\phi }{r_p}w^3\cos\delta\theta,\\
\chi_{r\theta}&=-(r_p-2)u_r u_r w^2\cos\delta\theta,\\
\chi_{r\phi}&=u_r u_\phi \cos^4\delta\theta+\frac{(r_p-3)u_r u_\phi }{(r_p-2)r_p}w^1\cos^4\delta\theta+\left(\frac{u_\phi u_\phi }{r_p}-(r_p-2)u_r u_r \right)w^3\cos^3\delta\theta,\\
\chi_{\theta\phi}&=-(r_p-2)u_r u_\phi w^2\cos^3\delta\theta,\\
\chi_{\phi\phi}&=-2(r_p-2)u_r u_\phi w^3\cos^5\delta\theta+u_\phi u_\phi \cos^4\delta\theta+\frac{2u_\phi u_\phi }{r_p}w^1\cos^4\delta\theta.
\end{split}
\end{equation}


%

\end{document}